\documentclass[twocolumn]{aastex63}
\usepackage{color, here}
\usepackage[version=4]{mhchem}
\usepackage{bm}
\usepackage{lineno}
\nolinenumbers
\received{}
\revised{}
\accepted{}
\submitjournal{ApJ}

\shorttitle{stellar evolution and disk fraction}
\shortauthors{Komaki and Yoshida}

\newcommand{\rg}{r_{\rm g}}

\newcommand{\cs}{c_{\rm s}}
\newcommand{\au}{{\rm \,au}}

\newcommand{\gram}{{\rm \, g}}
\newcommand{\Msun}{{\rm \, M_\odot}}
\newcommand{\Lsun}{{\rm \, L_\odot}}
\newcommand{\cm}{{\rm \, cm}}
\newcommand{\erg}{{\rm \, erg}}
\newcommand{\eV}{{\rm \, eV}}
\newcommand{\keV}{{\rm \, keV}}

\newcommand{\Kelvin}{{\rm \,K}}

\newcommand{\Myr}{{\,\rm Myr}}
\newcommand{\yr}{{\,\rm yr}}
\newcommand{\second}{{\rm \, s}}
\newcommand{\figref}[1]{Figure~\ref{#1}}
\newcommand{\tabref}[1]{Table~\ref{#1}}

\newcommand{\secref}[1]{Section~\ref{#1}}

\begin{document}

\title{The effect of stellar evolution on dispersal of protoplanetary disks: Disk fraction in star-forming regions}

\author[0000-0002-9995-5223]{Ayano Komaki}
\affiliation{Department of Physics, The University of Tokyo, 7-3-1 Hongo, Bunkyo, Tokyo 113-0033, Japan}
\email{ayano.komaki@phys.s.u-tokyo.ac.jp}

\author[0000-0001-7925-238X]{Naoki Yoshida}
\affiliation{Department of Physics, The University of
Tokyo, 7-3-1 Hongo, Bunkyo, Tokyo 113-0033, Japan}
\affiliation{Kavli Institute for the Physics and Mathematics of the Universe (WPI), UT Institute for Advanced Study, The University
of Tokyo, Kashiwa, Chiba 277-8583, Japan}
\affiliation{Research Center for the Early Universe (RESCEU), School of
Science, The University of Tokyo, 7-3-1 Hongo, Bunkyo, Tokyo 113-0033, Japan}



\begin{abstract}
We study the effect of stellar evolution on the dispersal of protoplanatary disks by performing one-dimensional simulations of long-term disk evolution.
Our simulations include viscous disk accretion, magnetohydrodynamic winds, and photoevaporation as important disk dispersal processes. 
We consider a wide range of stellar mass of $0.1$ -- $7M_{\odot}$, and incorporate the luminosity evolution of the central star. 
For solar-mass stars, stellar evolution delays the disk dispersal time as the FUV luminosity decreases toward the main sequence. 
In the case of intermediate-mass stars, the FUV luminosity increases significantly over a few million years, driving strong photoevaporation and enhancing disk mass loss during the later stages of disk evolution. This highlights the limitations of assuming a constant FUV luminosity throughout a simulation. 
Photoevaporation primarily impacts the outer regions of the disk
and is the dominant disk dispersal process in the late evolutionary stage.
Based on the results of a large set of simulations, we study the evolution of a population of star-disk systems and derive the disk fraction as a function of time. 
We demonstrate that the inclusion of stellar luminosity evolution can alter the disk fraction by several tens of percent, bringing the simulations into closer agreement with recent observations. 
We argue that it is important to include the stellar luminosity evolution in simulations of the long-term dispersal of protoplanetary disks.
\end{abstract}



\section{Introduction} \label{sec:intro}
Protoplanetary disks (PPDs) are thought to be the primary birthplace of planets. 
The physical conditions such as gas density and dust content
of PPDs may be key factors in planet formation, and
understanding the evolution of PPDs is important in order to study planet formation 
as well as the evolution of young stellar systems.
In particular, it is crucial to understand how and how quickly the PPD around 
a newly born star is dispersed.

Infrared observations of star-forming regions are often used
to estimate the fraction of stars that have surrounding disks.
The disk fraction as a function of time (age) can then be used to infer 
the characteristic dispersal time of PPDs.
Recent observations suggest that the disk fraction decreases rapidly and that the typical disk lifetime is a few to several million years \citep[e.g.,][]{Haisch:2001,Meyer:2007,Hernandez:2007,Mamajek:2009,Bayo:2012,Ribas:2014,Michel:2021}. While there still remain large uncertainties in the disk lifetime, the inferred short timescale
might pose a challenge to theoretical models of planet formation that invoke long-term processes such as accretion of gas and solid bodies onto planet embryos.

Theoretically, three major physical mechanisms of PPD dispersal
have been proposed: viscous accretion, magnetohydrodynamic (MHD) winds, and photoevaporation.
Photoevaporation is caused by high-energy photons such as far-ultraviolet (FUV; $6\eV \lesssim h\nu < 13.6\eV $), extreme-ultraviolet (EUV; $13.6\eV \lesssim h\nu \lesssim 100\eV$), and X-ray ($100\eV \lesssim h\nu \lesssim 10\keV$) emitted from the central star \citep[e.g.,][]{Hollenbach:1994,Shu:1994,Gorti:2009,Owen:2010,Tanaka:2013,Alexander:2014,Wang:2017,Nakatani:2018}.
The high-energy photons effectively heat the disk gas through photo-ionization of various elements and  photoelectric effect on dust grains.
Previous studies often consider EUV and X-ray radiation as the main sources of disk heating \citep{Owen:2010,Tanaka:2013,Picogna:2019},
but FUV photons can play a substantial role in photoevaporation by penetrating deep into the disk \citep{Gorti:2009}.
\cite{Komaki:2021} conducted two-dimensional radiation hydrodynamics simulations to show that the FUV radiation from the central star drives strong photoevaporative flows through photoelectric heating for a wide range of stellar masses.
In general, photoevaporation plays a crucial role in disk dispersal, especially in the late to the final stages of PPD evolution \citep{Clarke:2001,Alexander:2014}.

It is important to note that the luminosity of a pre-main sequence star varies significantly over a few million years.
Since the mass-loss rate due to disk photoevaporation roughly scales with stellar luminosity, stellar evolution may
need to be incorporated to address the dispersal process and the lifetime of PPDs.
\cite{Kunitomo:2021} performed one-dimensional long-term disk simulations with the effect of stellar evolution included. 
They showed that the FUV luminosity of an intermediate to massive star increases by several orders of magnitude before reaching the main-sequence phase.
The increased luminosity leads to stronger photoevaporation
and thus to shorter disk lifetimes around the stars.
Clearly, it is important to incorporate the stellar evolution
in the study of long-term PPD evolution.

In the present paper, we examine the effect of stellar evolution on the PPD dispersal considering viscous accretion, MHD winds, and photoevaporation. We largely follow the methodology of \cite{Komaki:2023} 
and perform a set of one-dimensional simulations.
Since disks are often observed in groups of stars for statistical analyses, 
it is essential to examine disk evolution and derive statistics for a stellar cluster.
To this end, we perform disk evolution simulations with varying stellar mass, disk mass, and disk radius to explore the effects of these initial parameters on the disk dispersal timescale.
We assume a star-forming region with a large number of star-disk systems and 
follow the disk evolution for each system.
We then analyze the collective behavior of a group of disks.
Finally, we compare our simulation results with recent observations 
and address the impact of stellar evolution on disk dispersal and the disk fraction.

The rest of the present paper is organized as follows.
We describe our numerical methods in Section 2.
The results of one-dimensional disk simulations and
those of population evolution are shown in Section 3.
We discuss the result in comparison with observations in Section 4. Summary and concluding remarks are given in Section 5.

\section{Numerical simulations}
We perform one-dimensional disk evolution simulations by incorporating viscous accretion, MHD winds, photoevaporation, and stellar evolution.
Here, we provide a summary of the key physical processes.
The details of the disk evolution calculation can be found in \cite{Komaki:2023}.

We follow the evolution of the gas surface density, $\Sigma$,
by solving the governing equation
\[
\begin{split}
&\frac{\partial\Sigma}{\partial t} + \frac{1}{r}\frac{\partial}{\partial r}\left(r\Sigma v_{r}\right) + \dot{\Sigma}_{\textrm{w}} + \dot{\Sigma}_{\textrm{pe}}=0,\\
&r\Sigma v_{r} = -\frac{2}{r\Omega}\left[ \frac{\partial}{\partial r}\left(r^2\Sigma\overline{\alpha_{r\phi}}\cs^2\right)+r^2\overline{\alpha_{\phi z}}\left(\rho\cs^2\right)_{\textrm{mid}} \right],
\end{split}
\]
where $\dot{\Sigma}_{\textrm{w}}$ and $\dot{\Sigma}_{\textrm{pe}}$ are the surface mass-loss rates due to MHD winds and photoevaporation, respectively.
In the equations, $v_{r}$, $\Omega$ and $\cs$ are the velocity in the $r$ direction, the Keplerian angular velocity, and the sound speed.
The subscript $\textrm{mid}$ refers to the value at the disk mid-plane.

The efficiency of viscous accretion and wind torque are given by $\overline{\alpha_{r\phi}}$ and $\overline{\alpha_{\phi z}}$, respectively.
Recent observations toward star-forming regions suggest that $\overline{\alpha_{r\phi}}$ varies by a few orders of magnitude in the range of $10^{-4}$--$10^{-2}$ \citep{Facchini:2017}.
We set $\overline{\alpha_{r\phi}}=10^{-4}$ for the disk around a $1\Msun$ star,
and assume that $\overline{\alpha_{r\phi}}$ is proportional to the stellar mass. 
Namely, the viscous accretion rate roughly scales with the stellar mass \citep{Muzerolle:2003,Calvet:2004}.

The mass loss due to MHD winds is calculated as 
\begin{equation}
\dot{\Sigma}_{\textrm{w}} = (\rho\cs)_{\textrm{mid}}C_{\textrm{w}}=\frac{\Sigma\Omega}{\sqrt{2\pi}}C_{\textrm{w}},
\label{eq:MHDwinds}
\end{equation}
where $C_{\textrm{w}}$ expresses a dimensionless mass flux 
related to the energy used for launching gas winds. \citet{Suzuki:2016} shows that
it is given by
\[
C_{\textrm{w}} = (1-\epsilon_{\textrm{rad}})\left[ \frac{3\sqrt{2\pi}\cs^2}{r^2\Omega^2}\overline{\alpha_{r\phi}}+\frac{2\cs}{r\Omega}\overline{\alpha_{\phi z}} \right].
\]
Here, $\epsilon_{\textrm{rad}}$ denotes the fraction of energy converted to radiation that contributes to heating of the gas, while the rest of the energy liberated by accretion is used to drive gas flows.

The disk gas temperature is determined by a combined effect of viscous heating and stellar irradiation \citep{Suzuki:2016}.
We calculate the gas temperature from the balance equation
\[
T^4 = T_{\rm irr}^4 + T_{\rm vis}^4.
\]
In our calculation, we conservatively assume that a fraction ($\epsilon_{\textrm{rad}}=0.9$) of the energy is used to heat the gas and dust in the disk.
The irradiation temperature, $T_{\textrm{irr}}$, is derived from the bolometric luminosity, $L_{\textrm{bol}}$, as
\[
T_{\textrm{irr}} = 280\Kelvin \left(\frac{L_{\textrm{bol}}}{\Lsun}\right)^{1/4}\left(\frac{r}{1\au}\right)^{-1/2}.
\]
We update $L_{\textrm{bol}}$ along with the stellar evolution over time.

We adopt the surface density mass-loss rates by photoevaporation
that are derived from the detailed numerical simulations of \citet{Komaki:2021} (K21 hereafter).
In K21, a set of two-dimensional photoevaporation simulations
are performed, which solve radiative transfer, hydrodynamics, and non-equilibrium chemistry in a dynamically coupled, self-consistent manner.
An additional set of simulations with different FUV luminosities are used to show that the mass loss rate increases with 
$L_{\textrm{FUV}}$ as $\dot{M}_{\textrm{pe}}\propto L_{\textrm{FUV}}^{0.55}$ to a good approximation.
To incorporate the variation in FUV luminosity due to stellar evolution, we define a scaling parameter, $\beta=(L_{\textrm{FUV}}/L_{\textrm{FUV,sim}})^{0.55}$, where $L_{\textrm{FUV,sim}}$ represents the FUV luminosity used in the hydrodynamics simulations.
In our one-dimensional simulations, the corresponding mass-loss rate by photoevaporation is calculated as
\[
\Sigma_{\textrm{pe}}=\beta\Sigma_{\textrm{pe,sim}}.
\]

The X-ray luminosity of an intermediate star decreases at the age of a few million years because it does not have a convective zone on the surface and therefore the magnetic activity is inhibited \citep{Flaccomio:2003}.
The X-ray luminosity decreases when the FUV luminosity begins to increase.  
Because $\dot{M}_{\textrm{pe}}$ is more correlated with the FUV luminosity, we only consider the time variation of the FUV luminosity in the simulations in the present paper.
We calculate the FUV luminosity of the central star at each time step to update $\Sigma_{\textrm{pe}}$.
We consider three mechanisms contributing to the FUV emitting process; accretion luminosity, photospheric emission, and chromospheric emission.
A fraction of the gravitational energy is released in the form of radiation when the gas is accreted onto the stellar surface.
We assume that $4\%$ of the gravitational energy is emitted as FUV photons \citep{Calvet:1998}.
The accretion rate is calculated directly in our simulation using 
\[
\dot{M}_{\textrm{acc}}=-2\pi(r v_r \Sigma).
\]
We calculate the accretion rate at the inner computational boundary, $r=0.01\au$.
The chromospheric component is given by $L_{\textrm{FUV,ch}}=10^{-3.3}L_{\textrm{bol}}$, where $L_{\textrm{bol}}$ expresses the bolometric luminosity \citep{Valenti:2003}.
We follow \cite{Kunitomo:2020} to calculate the photospheric FUV luminosity.
We use the stellar spectrum code PHOENIX, which calculates the stellar spectra considering more than 90 elements and a variable effective temperature \citep{Husser:2013}.
We integrate the spectrum to derive the ratio of photospheric FUV luminosity to bolometric luminosity, denoted as $\gamma=L_{\textrm{FUV,ph}}/L_{\textrm{bol}}$.
The photospheric component is calculated by multiplying $\gamma$ by $L_{\textrm{bol}}$ that is calculated by MESA \citep{Paxton:2011}.
Using the three components allows us to model accurately the FUV luminosity during stellar evolution.

In order to investigate the impact of stellar evolution on disk dispersal process quantitatively, we also perform simulations without considering stellar luminosity evolution.
In these reference runs, we assume that the luminosity of the central star and the resulting photoevaporation rate profile remain time-independent.
The adopted stellar luminosities are listed in \tabref{tab:luminositylist}.

\begin{table*}[]
\begin{tabular}{|c||c c c c c c c c c c|}\hline
    $M_{*}$ [$\Msun$] & 0.1 & 0.3 & 0.5 & 0.7 & 1.0 & 1.7 & 2.0 & 3.0 & 5.0 & 7.0 \\ \hline
    $L_{\textrm{bol}}$ [$\Lsun$] & 0.06 & 0.26 & 0.93 & 1.72 & 2.34 & 5.0 & 6.4 & 14.9 & 554.5 & 1687 \\ \hline
    log $L_{\textrm{FUV}}$ [$\erg \second^{-1}$] & 30.7 & 30.8 & 30.9 & 31.3 & 32.0 & 32.3 & 32.4 & 32.3 & 36.0 & 36.5 \\ \hline
    log $L_{\textrm{EUV}}$ [$\second^{-1}$] & 39.7 & 39.9 & 40.1 & 40.5 & 40.7 & 41.0 & 41.0 & 39.0 & 40.0 & 44.1 \\ \hline
    log $L_{\textrm{X-ray}}$ [$\erg\second^{-1}$] & 29.2 & 29.9 & 30.3 & 30.5 & 30.8 & 31.1 & 27.9 & 28.7 & 29.3 & 33.1 \\ \hline
\end{tabular}
\caption{The list of stellar mass, disk mass, and disk radius we used for the simulations.}
\label{tab:luminositylist}
\end{table*}

In star-forming regions, a population of star-disk systems is observed, and the disk fraction is used as an indicator of the evolutionary stage of the population.
For close comparison with such observational data, we perform disk evolution simulations for a population of stars.
We consider basic physical parameters such as stellar mass, disk mass, and disk radius to represent the diversity of disks.
We perform a large set of one-dimensional disk evolution simulations for various star-disk systems to reproduce the diverse evolutionary patterns.
Specifically, we vary the stellar mass, disk mass, and disk radius and configure 250 different combinations of these quantities to cover the diverse environment of young stars.
The set of parameters we use is shown in \tabref{tab:parameterlist}.
\begin{table}[]
\centering
\begin{tabular}{|c|c|}\hline
    $M_{*}$ & 0.1, 0.3, 0.5, 0.7, 1.0, 1.7, 2.0, 3.0, 5.0, 7.0 $\Msun$ \\ \hline
    $M_{\textrm{disk}}$ & 0.1, 0.3, 1.0, 3.0, 10 $\times 0.0117M_{*}$ \\ \hline
    $R_{\textrm{disk}}$ & 0.3, 0.7, 1.0, 1.3, 1.7 $\times (M_{*}/1\Msun)30\au$ \\ \hline
\end{tabular}
\caption{The list of stellar mass, disk mass, and disk radius we used for the simulations.}
\label{tab:parameterlist}
\end{table}

For each system, we estimate the disk dispersal timescale, $t_{\textrm{dis}}$, defined as the period during which the disk remains observable in the near-infrared spectrum. Following the criteria of \cite{Kimura:2016}, we assume that small dust grains are observable when the surface density satisfies $\Sigma_{\textrm{dust}} \geq 0.1 \gram\cm^{-2}$ and the temperature is at least $300 \Kelvin$.
The critical dust surface density is chosen to ensure that the local optical depth reaches unity. For small dust grains with size $\leq 10 \,\mu \rm{m}$, the absorption coefficient is approximately $10 \cm^{2}\gram^{-1}$, justifying our criterion of $\Sigma_{\textrm{dust}} \geq 0.1 \gram\cm^{-2}$.
Although our simulations primarily follow gas evolution, near-infrared observations probe the presence of dust in protoplanetary disks. Throughout this study, we assume a spatially constant dust-to-gas mass ratio, $\mathcal{D}$, fixed at 0.03. 
This ratio is motivated by \cite{Gorti:2015}, who modeled the co-evolution of gas and dust. According to their results at $t = 1.5 \Myr$, the dust-to-gas mass ratio is approximately 0.03 at $r = 30 \au$, which is the typical survival radius of the gas disk in our simulations. The criterion of $\Sigma_{\textrm{dust}} \geq 0.1 \gram\cm^{-2}$ corresponds to a gas surface density of $\Sigma \geq 3.3 \gram\cm^{-2}$.

To create a realistic stellar "cluster", we generate a total of 10,000 star-disk systems by Monte-Carlo sampling of the physical parameters.
Instead of running simulations for each of the 10,000 systems, we interpolate the results from our 250 direct simulations to estimate the disk dispersal time.
The disk fraction is then calculated by counting the number of surviving disks among the 10,000 disks as a funct
ion of time.
In order to test the accuracy of the Monte-Carlo sampling and interpolation, we randomly selected four sets of disk parameters of $(\,M_{*},\tilde{M}_{\textrm{disk}},\tilde{R}_{\textrm{disk}})=(0.3,0.2,0.7),(1.0,0.7,1.0),(1.0,7.0,1.5),(3.0,0.6,1.0)$.
The units of $\tilde{M}_{\textrm{disk}}$ and $\tilde{R}_{\textrm{disk}}$ are $0.0117M_{*}$ and $30\au(M_{*}/1\Msun)$ respectively.
We then compared the disk dispersal timescales obtained from direct simulations with those estimated through the interpolation as described in the above. The differences in the dispersal times were found to be less than 0.65 Myr, which corresponds to 6.6\% of the disk dispersal timescale.

We consider a wide range of stellar mass of $0.1$--$7\Msun$, and adopt the initial mass function(IMF) suggested by \cite{Kroupa:2001}, which is given by
\[
\begin{split}
\xi(m) &\propto m^{-\alpha}\\
& \alpha = 1.3\,(0.1\Msun<m\leq0.5\Msun)\\
& \alpha = 2.3\,(0.5\Msun<m).
\end{split}
\]
We consider this range because $0.1\Msun$ is the minimum mass limit of observations \citep{Michel:2021}.
We also exclude stars larger than $7\Msun$, because they are relatively rare and their disks tend to disperse rapidly because of the intense radiation of the host star.
The range of stellar mass allows us to focus on the typical population of an observed star-forming region and to compare our results with observations.

\begin{figure*}
       \centering
         \includegraphics[width=\linewidth,clip]{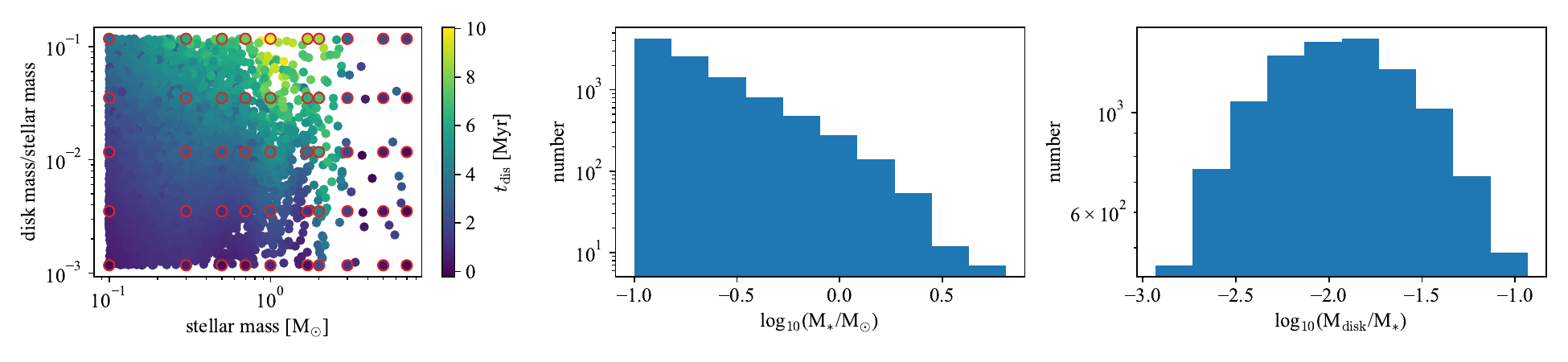}
         \caption{Physical quantities used in our simulations. The left panel shows the resulting lifetimes of simulations with various stellar mass and disk mass.
         The red circles indicate the lifetimes calculated directly for 250 runs. The color scale shows the lifetimes for a sample of 10,000 disks, interpolated from the results of the 250 runs.
         The other two panels on the right show the distribution of the stellar mass and disk mass for the 10,000 star-disk systems in our simulation. 
         }
         \label{initialparam}
\end{figure*}

The disk mass, $M_{\textrm{disk}}$, is varied in a range of $0.0117\times0.1$--$10 ~M_{*}$ with $M_{*}$ denoting the host star mass (see \figref{initialparam}).
Observations towards T Tauri stars suggest a proportional relation between stellar mass and disk mass \citep{Ansdell:2016,Andrews:2020}, while there is also a broad dispersion in disk mass over a few magnitudes \citep{RuizRodriguez:2018}.
We assume that the ratio of disk mass to stellar mass is
given by a log-normal Gaussian distribution with a mean ratio of $M_{\textrm{disk}}/M_{*}=0.0117$ that is given by the minimum solar mass model of \cite{Hayashi:1981}.
The dispersion is set to $\sigma=0.35$ dex \citep{Tychoniec:2018}.

The initial gas density distribution is given by
\[
\Sigma=\Sigma_{1\au}\left(\frac{r}{1\au}\right)^{-3/2}\exp{\left(-\frac{r}{r_{\textrm{disk}}}\right)},
\]
where $R_{\textrm{disk}}$ denotes the disk cut-off radius.
Recent ALMA observations of star-forming regions indicate a correlation between disk radius and stellar mass \citep{Eisner:2018,Andrews:2018}, while others find no clear correlation \citep{Hendler:2020}.
It may be reasonable to set the disk radius to be proportional to stellar mass. We thus assume the mean radius of $R_{\textrm{disk}}=30\au(M_{*}/1\Msun)$,
and assume a log-normal Gaussian distribution around the mean with dispersion $\sigma=0.1$ dex \citep{Tobin:2020}.

\figref{initialparam} shows the distribution of the stellar mass, disk mass, and disk radius for the 10,000 disks considered in our calculation.
We reproduce the variety of PPDs in a typical star-forming region with these initial parameters. We compare the simulation result with the observed disk fraction.

For individual systems, we continue the calculation until the disk gas mass decreases to $M_{\textrm{disk}}<10^{-10}\Msun$ or the time reaches $100\Myr$.
The end time is not significantly affected by the threshold mass because the disk gas disperses rapidly in the final stage of evolution.


\section{Results}   \label{sec:results}
In this section, we present the results of our simulations.
We first discuss the effect of stellar evolution on disk dispersal in \secref{sec:result1}.
We then study the disk fraction and its time variation of a star-forming region in \secref{sec:result2}.

\subsection{The impact of stellar evolution}\label{sec:result1}
\begin{figure}
       \centering
         \includegraphics[width=\linewidth,clip]{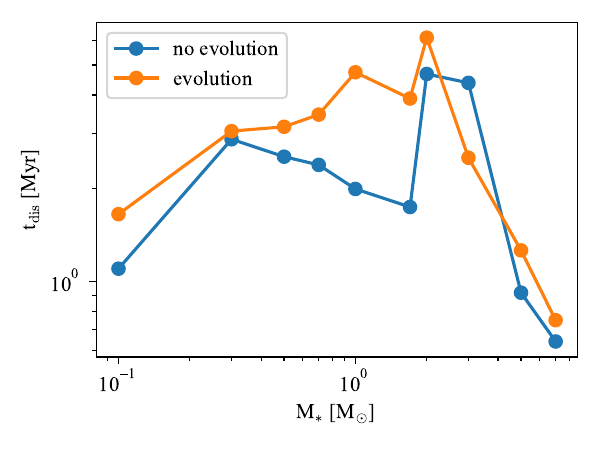}
         \caption{The disk dispersal time as a function of stellar mass. The blue and orange lines represent $t_{\textrm{dis}}$ by calculations with and without stellar evolution, respectively.}
         \label{lifetime}
\end{figure}
\figref{lifetime} shows the disk dispersal time as
defined in the previous section for the runs with the fiducial disk mass and radius.
The dispersal timescale varies significantly from a few to several million years depending on the stellar mass.

The long-term evolution of the disk is summarized as follows (see also \citet{Komaki:2023}). 
Initially, the disk mass loss is driven by MHD winds in the inner disk, but MHD winds weaken over time as the gas surface density decreases because $\dot{\Sigma}_{\textrm{w}}$ is proportional to the surface density.
In contrast, photoevaporation is the major dispersal process in the outer disk with $r>\rg$.
As the MHD winds weaken, photoevaporation becomes the main process responsible for the total gas dispersal throughout the disk.

\figref{lifetime} shows that the disk dispersal timescale has a peak around $2$--$3\Msun$ if stellar evolution is not included. 
This is consistent with the simulation of the "low-viscosity" case in \cite{Ronco:2023}.
Intermediate-mass stars do not develop a convective zone near the surface, and thus the emissivity of high-energy photons is reduced, which generally leads to low disk mass-loss rates due to photoevaporation.
The effect of the stellar activity is significant enough to extend the overall disk dispersal timescale for intermediate-mass stars.

When stellar evolution is incorporated into the calculation, the disk dispersal timescale shows different trends depending on the stellar mass.
In the range of $0.5\Msun\leq M_{*}\leq2\Msun$, the disk dispersal time is longer compared to the case without stellar evolution.
During the pre-main sequence phase, the low-mass stars
evolve on the Hayashi track, and the bolometric luminosity decreases. For a $1\Msun$ star, this phase lasts for several million years.
By the time photoevaporation becomes the dominant dispersal process, $\dot{\Sigma}_{\rm pe}$ decreases to half of the initial value.
Since we assume the luminosity at the age of $1\Myr$ and the accretion rate of $\dot{M_{\textrm{acc}}}=1\times10^{-8}(M_{*}/\Msun)\Msun\yr^{-1}$ in the calculation without stellar evolution, 
the time variation of the FUV luminosity relative to this value affects the efficiency of photoevaporation.
This results in longer dispersal times for the low-mass stars when stellar evolution is incorporated.
For lower-mass stars, the disk dispersal timescale is shorter because the survival criterion requires $T\geq300\Kelvin$. 
Low-mass stars have lower luminosities, which further decrease as the star moves along the Hayashi track. 
As a result, the region where the dust disk can be observed shrinks.
In contrast, a $2\Msun$ star reaches the main-sequence during the final stages of disk dispersal.
The radius of the $T\geq300\Kelvin$ disk region increases by a factor of 1.5 in the last $1\Myr$, leading to longer disk lifetimes.
This results in shorter disk lifetimes as stellar mass decreases, with a peak in the dispersal timescale around $2\Msun$ stars.


Stars with masses greater than $3\Msun$ reach the main-sequence phase within a few million years.
Both the bolometric luminosity and effective temperature increase before the disk is fully dispersed.
\begin{figure}
       \centering
         \includegraphics[width=\linewidth,clip]{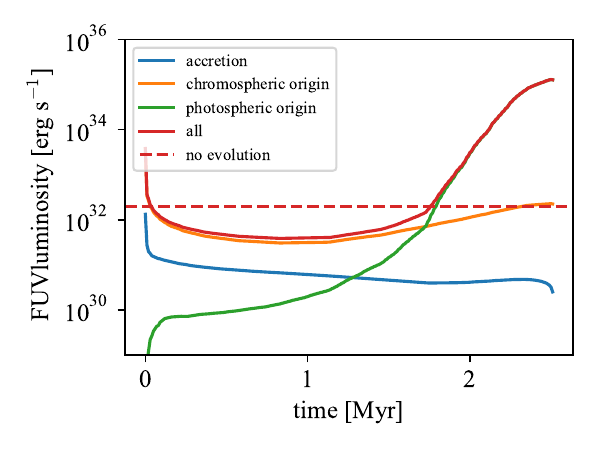}
         \caption{The evolution of FUV luminosity from a $3\Msun$ star. The blue, orange, and green lines represent accretion origin, chromospheric origin, and photospheric origin. The red line shows a sum of FUV luminosity by all three mechanisms, and the red dotted line represents the FUV luminosity assumed in calculation without stellar evolution.}
         \label{FUV3Msun}
\end{figure}
\figref{FUV3Msun} shows the evolution of FUV luminosity for a $3\Msun$ star.
The FUV radiation from the stellar chromosphere is the main source in the first few million years.
When the star reaches the main-sequence, the photospheric emission becomes more significant because of the increase in both bolometric luminosity and effective temperature.
When the effective temperature increases to $30,000\Kelvin$, the contribution of the FUV photons becomes dominant over the whole stellar spectrum.

The mass-loss rate by photoevaporation increases with the FUV luminosity, following the relation $\dot{M}_{\textrm{pe}}\propto L_{\textrm{FUV}}^{0.55}$.
This relationship results in a significant increase in the mass-loss rate, shortening the disk dispersal timescale by approximately 1.9 million years compared to a calculation that assumes a constant FUV luminosity fixed at the value at the age of 1 Myr. 
The assumed constant FUV luminosity is overestimated due to the high accretion rate of $3.0\times10^{-8}\Msun\yr^{-1}$ as shown in \figref{FUV3Msun}.
Nevertheless, since the FUV luminosity reaches its minimum at around 1 Myr, incorporating stellar evolution introduces a substantial difference in disk dispersal timescale.


\subsection{Disk fraction and comparison to observations}\label{sec:result2}

\begin{figure}
       \centering
         \includegraphics[width=\linewidth,clip]{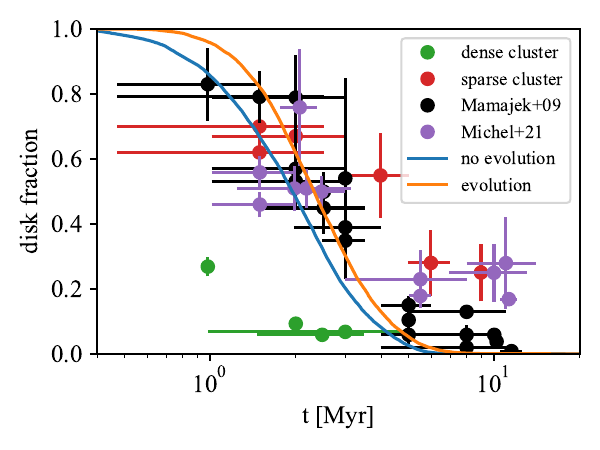}
         \caption{The disk fraction evolution over time. The blue and orange lines represent the results of calculations with and without stellar evolution. The black plots represent the observed disk fraction obtained by \cite{Pfalzner:2022}.}
         \label{diskfraction}
\end{figure}
\figref{diskfraction} compares the disk fraction over time with and without stellar evolution. A number of recent observations are also plotted there.
Our simulations show that the disk fraction decreases rapidly within $\sim3\Myr$.
The derived disk fraction largely reflects the survival rate of disks around low-mass stars. This is attributed to the assumption that the stellar mass distribution follows the IMF with a strong decline towards massive end.

As \figref{lifetime} shows, $t_{\textrm{dis}}\sim2\Myr$ for a $0.1\Msun$ star, which largely sets the time when the disk fraction of the population drops significantly.
The disk fraction with stellar evolution is systematically higher by about $20\%$ at the age of $3\Myr$ compared to the case without stellar evolution.
This is owing to the relatively long dispersal timescale of disks around $\sim 1\Msun$ stars.
The trend highlights the importance of including stellar evolution in the present study.
The overall effect of stellar evolution is evident in the disk fraction, and the differences in dispersal timescales are most pronounced for solar to intermediate-mass stars.
We further discuss the implication in comparison to the observations in \secref{discussion1}.

Although most of the disks in our sample disperse in several million years, some disks survive even at $t > 6\Myr$. The long-lived disks may be important
birthplaces of planets, and thus we study the characteristics in detail. 
We perform a set of disk evolution simulations varying the initial disk mass in the range of 0.00117--0.117 times the stellar mass to understand in particular the dispersal process of massive disks.
We have found that the dependence of $t_{\textrm{dis}}$ on the initial disk mass varies with the stellar mass. 
Specifically, the disk dispersal timescale scales as $\propto M_{\textrm{disk,ini}}^{0.40}$ for a $1\Msun$ star and $\propto M_{\textrm{disk,ini}}^{0.17}$ for a $3\Msun$ star.
Although a large disk mass extends the disk lifetime, the longer dispersal time also allows enhanced photoevaporation driven by the increasing luminosity of a $3 \Msun$ star.
As a result, the disk dispersal timescale for a $3 \Msun$ star does not increase as significantly as it does for a $1 \Msun$ star.
This result highlights the influence of stellar evolution for intermediate-mass stars.  Recent observations suggest a wide variation of the initial disk mass, and thus it is important to notice that the effect of increasing stellar luminosity is more pronounced for a more massive disk.

While massive disks are dispersed quickly around high-mass stars, it takes more time around intermediate-mass stars
whose FUV luminosity is not as strong as dispersing the correspondingly massive disks quickly.
This highlights that the initial disk mass and the evolution
of the stellar luminosity non-trivially determine the lifetime of the disk. Clearly, detailed simulations such as we explore here are needed for quantitative studies.


\cite{Weder:2023} performed one-dimensional simulations of disk evolution
by incorporating viscous accretion, MHD winds, internal photoevaporation by EUV photons, and external photoevaporation by FUV photons.
In our simulations, we employ the surface mass-loss rate derived from 
radiation-hydrodynamics simulations that incorporate EUV, FUV, X-ray photons from the central star.
If we assume that the FUV luminosity is $L_{\textrm{FUV}}=10^{32}\erg\second^{-1}$, corresponding to an FUV flux of $2.2\times10^{5}\,G_{0}$ at $r=100\au$.
This is close to the strong FUV field scenario presented by \cite{Weder:2023},
and the simulations results are consistent overall.
The disk fraction decreases to half within approximately $2\Myr$, and is lower than the observed disk fraction \citep{Michel:2021}.
However, we have found a $\sim 20\%$ increase in the disk fraction by including stellar evolution in our calculations, which brings the disk fraction closer to the observational data.

Note that the observational estimate of the age of a cluster depends on the assumed stellar evolutionary model. 
Recent theoretical studies suggest that cool stars evolve 
on low-temperature and high-luminosity tracks
if inefficient convection caused by magnetic fields is considered \citep[e.g.,][]{Feiden:2016}. 
This may lead to a higher disk fraction at a given age \citep{Richert:2018}. 
Although it is still difficult to estimate the absolute disk lifetime accurately, we argue that including stellar evolution aligns the disk fraction more closely with recent observations.

\section{Discussion} 

In this section, we discuss uncertainties that could affect our results
and also the implications for observations.

\subsection{Model Uncertainties}
\label{sec:models}

We have seen a clear variation of disk dispersal time as a function of stellar mass (Figure 2).
It is important to note that the criterion to judge when a disk is dispersed depends on the stellar mass (luminosity). 
Specifically, it takes several million years for the disk gas around low-mass stars to disperse, but the dust grains remain {\it observable} only for $1$--$4\Myr$ because of the low bolometric luminosity of low-mass stars.
In our model, both the gas and dust temperatures are determined
by a combined effect of irradiation from the central star and viscous heating. The temperatures are roughly proportional to $\propto L_{\textrm{bol}}^{1/4}$. 
The resulting dust temperature is low for low-mass stars, 
which yields the observable dust grains to be located only within the innermost region of $r<0.3\au$.
Therefore, the "observable" time for dust around low-mass stars is short compared to those around more massive stars.

Observations of T Tauri stars have revealed significant diversity in disk masses, spanning several orders of magnitude. Some studies suggest a proportional relationship between stellar mass and disk mass \citep{Ansdell:2016,Andrews:2020}, while others do not find such clear trend \citep{RuizRodriguez:2018}.
Yet other observations suggest that the relationship between disk mass and stellar mass could be as steep as $M_{\textrm{disk}}\propto M_{*}^{1.8}$ \citep{Pascucci:2016,Manara:2023}. In light of this (uncertainty), we conducted disk evolution simulations using the steep relation while keeping the disk mass around a $1\Msun$ star constant.

The additional simulations show that the disk dispersal timescale, $t_{\textrm{dis}}$, scales with the initial disk mass in all the cases. We have also found that the impact of stellar evolution on disk dispersal depends on the initial disk mass. For massive stars, even massive disks are dispersed rapidly due to intense photoevaporation, while for low-mass stars $t_{\textrm{dis}}$ is smaller for lower disk mass.
Hence the steep relationship between disk mass and stellar mass emphasizes the dependence of $t_{\textrm{dis}}$ on stellar mass. 

Let us discuss a specific case with $M_* = 3\Msun$ in detail. When the initial disk mass is set to $M_{\textrm{disk}} = 0.0117M_{*}$, the disk disperses shortly after the star reaches the main sequence, which means that stellar luminosity evolution has minimal impact on disk evolution. However, when the initial disk mass is increased ten times to $M_{\textrm{disk}} = 0.117M_{*}$, the dispersal time is reduced by half compared to the case without considering stellar evolution. This is because the disk is exposed to enhanced irradiation from the main sequence star for a longer period.
These complicated effect highlights not only the importance of stellar evolution for disk dispersal but also that the initial disk mass is a key quantity that determines the disk lifetime.


An important element in our simulations is that the so-called weak wind case is 
adopted as our fiducial model. 
There are a variety of possible combinations of physical processes such as strong/weak winds, the strength of magnetic fields.
It remains uncertain which model best describes the actual disk evolution. 
Also the "best" model 
(combination) may likely be different for different types of star/disk systems. In our present study, it would probably be important to evaluate the impact of stellar evolution across different models.

To this end, we performed additional disk evolution simulations for both the strong wind case and the no-magnetic-fields case.
In the strong wind model, the disk dispersal timescale is short with $t_{\textrm{dis}} < 3\Myr$ for all stellar mass cases, and it peaks around $3\Msun$. The influence of stellar evolution is limited because the inner disk dispersal is primarily driven by MHD winds, which account for more than 50\% of the disk's dispersal. As a result, the disk dispersal timescale generally decreases virtually in all the cases because strong MHD winds efficiently remove gas from the inner regions.

Our no-magnetic-fields case assumes that MHD winds are not launched. 
However, we still assume the fiducial parameter of $\overline{\alpha_{r\phi}}=10^{-4}$ that approximates a minimal effect of the turbulence viscosity in a disk.

In the no-magnetic-fields case, the relative contribution of photoevaporation increases, making stellar luminosity evolution to be more important. 
In this case, the disk dispersal timescale is longer for all stellar masses than in our fiducial case.  Also the dispersal time decreases roughly monotonically as the stellar mass increases. The most pronounced effect is observed in low-mass stars around $0.1\Msun$. Without strong influence of MHD winds, the gas remains in the disk for an extended period. The impact of stellar evolution is particularly notable for $3\Msun$ and lower-mass stars, confirming that assuming a constant stellar luminosity results in inaccurate estimates of the disk lifetime for the mass ranges.
Considering recent observations that suggest long disk lifetimes, we conclude that it is necessary to include stellar evolution in models of long-term disk evolution.

\subsection{Observed disk fraction}
\label{discussion1}
We compare our results with the observations presented by \cite{Pfalzner:2022}, who compiled a set of data from various star-forming regions and examined how the disk fraction varies among them.
\cite{Pfalzner:2022} categorized the observed systems into dense and sparse clusters and examined the disk fractions in the two groups.
As shown in \figref{diskfraction}, our simulation predicts
significantly larger fractions than those for dense clusters.
This is expected because our model does not incorporate external photoevaporation and thus approximates the physical conditions in sparse clusters.
Dense clusters experience higher field FUV flux from nearby massive stars, which enhances mass loss due to external photoevaporation and consequently leads to shorter disk lifetimes \citep{Adams:2004,Anderson:2013,Haworth:2019}.

We also compared our results with observations by \cite{Michel:2021}, who selected Class III disks at various ages and calculated the disk fraction for more than 10 systems.
Their focus was primarily on the nearby star-forming regions, mostly within $200$ pc, for which the observational selection bias is well understood, and also
the population includes faint, low-mass stars.
This makes the comparison with our theoretical model more straightforward.
Our simulation predicts the time evolution of the disk fraction, consistent with the observations in the early stages, roughly within a few million years.
However, \cite{Michel:2021} note that structured disks with dust traps
can cause observable timescales to exceed $8\Myr$,
which appears to be longer than our simulation result.
Interestingly, \cite{Michel:2021} also argue that the existence of dust traps prevents dust grains from being accreted rapidly onto the star.
As a result, the dust mass may not decrease as it would in a disk without any substructure.

We note that our model assumes a spatially constant dust-to-gas mass ratio, leading to a uniformly smooth distribution of dust grains, similar to the gas distribution.
This assumption might result in a lower calculated disk fraction compared to real scenarios with dust traps.

Our model assumes that the stellar mass distribution follows the designated IMF, but the actual stellar mass distribution might not match the assumed one.
\cite{Michel:2021} analyzed stars ranging from M5 to F0 spectral types, and found that the mass distribution of the stars does not closely match the standard IMF. Notably, the number of low-mass stars is significantly small.
Because the disk dispersal time decreases
substantially to $\sim 3\Myr$ at the low-mass end of $0.1 M_{\odot}$, the existence or the relative fraction of low-mass stars within a
star-forming region critically determines the disk fraction of the system at a given age.
For example, if the initial stellar mass distribution were adjusted to match the observed distribution, the calculated disk fraction could be higher and could become more consistent with the observation.
Motivated by this, we have performed a simple test calculation by assuming that the stellar mass distribution follows exactly that of the Class II stars in Lupus \citep{Ansdell:2018}.
The simulations show that the disk fraction at $t=2\Myr$ increases by $\sim15\%$ compared to the result of our fiducial model with the Kroupa IMF.
Clearly, it is important to use realistic distributions of physical parameters for more accurate simulations and a fair,
direct comparison.

\subsection{stellar mass dependence}
Infrared observations suggest that disk lifetime varies with stellar mass \citep{Carpenter:2006,Yasui:2014,Ribas:2015}.
Specifically, disks around massive stars disperse rapidly, as indicated by lower disk fractions.
Similarly, \cite{Pfalzner:2022} reached a comparable conclusion using a different analytical approach, focusing on disk fractions in distant star-forming regions.
They noted that observations of star-forming regions more than 2000 pc away tend to reflect disk fractions around massive stars because of the low luminosity of low-mass stars.
The study concluded that the typical disk dispersal timescale for a massive star is $2$--$4\Myr$.

\begin{figure}
       \centering
         \includegraphics[width=\linewidth,clip]{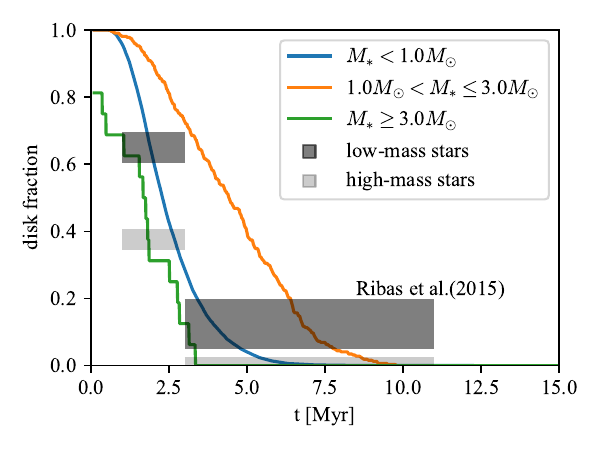}
         \caption{The blue, orange, and green lines represent the evolution of disk fraction for groups of $M_{*}<1\Msun$, $1\Msun\leq M_{*}\leq 3\Msun$, and $M_{*}>3\Msun$, respectively. The dark and light black lines denote the observed disk fraction of low-mass stars($M_{*}<2\Msun$) and high-mass stars($M_{*}\geq2\Msun$) by \cite{Ribas:2015}.}
         \label{diskfraction_mass}
\end{figure}
\figref{diskfraction_mass} shows the disk fraction of three groups:  $M_{*}<1\Msun$, $1\Msun\leq M_{*}\leq 3\Msun$, and $M_{*}>3\Msun$, corresponding to low-mass, intermediate-mass, and massive stars, respectively, out of 10000 disks in our calculation.
This motivates us to further examine the stellar mass dependence of disk lifetime.
Interestingly, the disk fraction of massive stars decreases rapidly within $4\Myr$, consistent with the observed trend in distant star-forming regions.
This decrease is attributed to strong photoevaporation caused by the intense radiation from the central star.
Disks around intermediate-mass stars have the longest lifetimes and the highest disk fractions.
In general, the longer time allows more time for planet formation, which could explain the occurrence rate of gas giants around intermediate-mass stars \cite{Ronco:2023}. 

Because low-mass stars typically dominate the population in number, it is important to examine their relative contribution to the observed disk fraction.
Our calculations show a typical dispersal time of $3$--$5\Myr$ for low-mass stars,
which is slightly shorter than suggested by recent observations. 
Our simulations follow the structure and evolution of gas in a PPD, and the dust distribution is derived using an assumed dust-to-gas mass ratio. 
Ideally, more detailed models are needed to treat the gas and dust evolution separately but in a coupled manner.
Given that low-mass stars have long gas lifetimes, often $\sim10\Myr$, the dust disk lifetime in the stars is more affected by the dust-to-gas mass ratio.

\figref{diskfraction_mass} shows the observational result from \cite{Ribas:2015}. They observed 11 star-forming regions and classified the member stars according to their age and mass, and considered $3\Myr$ and $2\Msun$ as characteristic thresholds. 
The result as shown in \figref{diskfraction_mass} suggests that the disk dispersal timescale is shorter for high mass stars
although with still large statistical uncertainty.

Our simulations show that the disk dispersal timescale peaks around $2\Msun$. This is the value adopted in \cite{Ribas:2015} as the boundary for low-mass/high-mass stars. Hence the exact stellar mass distribution influence the estimated disk fraction in the two groups.

We note that the disk dispersal timescale for low-mass stars is sensitive to the chosen disk evolutionary model as discussed in \secref{sec:models}. 
For example, in the no-magnetic-fields case, the disk may not disperse even within $100\Myr$, whereas in the weak MHD wind case, it disperses within $2\Myr$. 
This demonstrates that the estimated disk fraction can vary depending on the assumed model or the strength of magnetic fields.


\subsection{Dust Evolution}
There remains large uncertainties in the dust distribution within a disk.
We assumed a constant dust-to-gas mass ratio, which may not accurately reflect the differences in the gas and dust distribution.
Theoretical studies suggest that small dust grains are susceptible to entrainment by gas outflows \citep{Hutchison:2016}.
In our model, MHD winds could also entrain dust grains from the disk, which would then affect the mass-loss rate by photoevaporation.
Since the disk gas is mainly heated by FUV photons via photoelectric heating, the depletion of small dust grains can decrease the effect of photoevaporation.
However, \cite{Komaki:2024} show that the mass-loss rate by photoevaporation does not vary significantly in the range of $10^{-3}\leq\mathcal{D}\leq10^{-1}$.
We thus expect that the disk dispersal time is not drastically affected by entrainment, unless almost all of the small dust grains are lost.

Infrared and millimeter observations of PPDs suggest that gas and dust are distributed differently, but also the dust size distribution can vary significantly, possibly due to processes such as radial drift, coagulation, and settling \citep{DeGregorioMonsalvo:2013,Ansdell:2018,Grant:2018}.
Typically, the distribution of small dust grains is more extended than that of large grains.
By integrating these processes into our simulations, we will be able to compare our results with infrared observational data in more detail and provide more accurate predictions for disk dispersal timescales.

Observations of low-mass stars suggest that the abundance of PAH (polycyclic aromatic hydrocarbon) is only about 10
percent compared to other stars. Since the photoelectric effect is driven by  dust grains smaller than 100 \AA, the reduced PAH abundance could affect the efficiency of photoevaporation. \cite{Komaki:2023} showed that the mass-loss rate remains consistent within the range of $10^{-3} \leq \mathcal{D} \leq 10^{-1}$, but decreases significantly by a factor of 100 when $\mathcal{D}$ decreases to $10^{-4}$, compared to the case of $\mathcal{D} = 10^{-2}$.

To investigate whether/how the low abundance of PAH affects overall disk evolution and lifetime, we performed disk evolution simulations with a reduced surface mass-loss rate. We run the additional simulations with the stellar mass of 0.1--$1.0\Msun$. 
We estimate the disk dispersal timescale in the same manner as in fiducial case.
The disk dispersal timescale increased in all the cases, but the largest increase is found to be only by a factor of 1.37. Also the dependence of $t_{\textrm{dis}}$ on stellar mass remained unchanged. This is because photoevaporation primarily removes gas from the outer regions, whereas $t_{\textrm{dis}}$ reflects the observable timescale of the innermost regions.

\subsection{Other effects}
Several physical processes have been proposed as major machanism(s) of gas dispersal in protoplanetary disks, including accretion, MHD winds, and photoevaporation. Previous studies largely focused on understanding how each process works individually. In reality, these processes likely work in combination, and also mutual interaction between them would be important. For example, recent studies suggest that winds launched from the inner disk may partially shield radiation from the central star, reducing the mass-loss by photoevaporation \citep{Pascucci:2020, Pascucci:2023, Weder:2023}.

We study the effect by incorporating wind shielding in the following approximate manner. We turn off photoevaporation when the column density within 1 au exceed $N_{\rm H2,c} = 10^{20}\cm^{-2}$. This threshold is motivated by the photoevaporation simulations of \cite{Komaki:2021}, which show that the base of the photoevaporative flows corresponds approximately to $N_{\rm H2,c}$.

The test simulations show that the estimated disk lifetime remain unchanged in the case of $M_{*} = 1\Msun$. This is again explained by the same reason as in the previous section. Photoevaporation primarily affects the outer disk, while the disk dispersal timescale, as determined by observation in near-infrared, is more dependent on the inner disk regions.

\section{summary}
Infrared observations toward star-forming regions suggest that the decreasing disk fraction as a function of the system's age reflects a characteristic disk dispersal time, which has been estimated to be a few
to several million years.
Dispersal of protoplanetary disks is a highly nonlinear process that involves viscous disk accretion, MHD winds, photoevaporation, and stellar evolution and thus typically one needs to perform detailed numerical simulations to follow the long-term evolution of PPDs.
In this study, we integrate all these physical processes to represent realistic disk evolution and examine in particular the impact of stellar luminosity evolution on disk dispersal.
We have shown a strong impact of stellar evolution to stars of solar to intermediate mass. 
While photoevaporation primarily affects the outer regions of a disk and thus has a limited effect on the overall disk dispersal timescale, a dynamic change such as a stellar transition to the main sequence can alter the dispersal timescale through enhanced photoevaporation.
Stellar luminosity evolution plays a crucial role in the disk photoevaporation process, because a star reaches the main sequence on a similar timescale to the disk lifetime.
For a star with $M_{*} = 3\Msun$, the disk dispersal time is shortened because of an increase in the photoevaporation rate as the star reaches the main sequence. 
For stars with masses of $1$--$2\Msun$, using a constant FUV luminosity throughout the calculation fails to appropriately trace the disk dispersal timescale. This is because the bolometric luminosity decreases during the Hayashi phase, reducing the efficiency of photoevaporation.
As a result, the disk fraction for a population of star-disk systems differs by a few tens of percent compared to cases without stellar luminosity evolution.
We therefore suggest that incorporating stellar evolution is essential for accurately determining the longevity and characteristics of protoplanetary disks around intermediate-mass stars.

In future work, we will further explore the long-term evolution of PPDs by incorporating the formation, growth, and destruction of dust grains.
We will then be able to reproduce the distribution of small dust grains within a disk, to be compared with observations in detail.

\acknowledgments
We thank Hiroto Mitani and Takeru Suzuki for insightful discussions on stellar evolution.
This work was supported by JSPS KAKENHI Grant Number 23KJ0459. 
NY acknowledges financial support from JSPS grant 20H05847. Numerical computations were carried out on Cray XC50 at the Center for Computational Astrophysics, National Astronomical Observatory of Japan.

%
\vspace{5mm}





\bibliography{bibliography}

\begin{thebibliography}{}
\expandafter\ifx\csname natexlab\endcsname\relax\def\natexlab#1{#1}\fi
\providecommand{\url}[1]{\href{#1}{#1}}
\providecommand{\dodoi}[1]{doi:~\href{http://doi.org/#1}{\nolinkurl{#1}}}
\providecommand{\doeprint}[1]{\href{http://ascl.net/#1}{\nolinkurl{http://ascl.net/#1}}}
\providecommand{\doarXiv}[1]{\href{https://arxiv.org/abs/#1}{\nolinkurl{https://arxiv.org/abs/#1}}}

\bibitem[{{Adams} {et~al.}(2004){Adams}, {Hollenbach}, {Laughlin}, \& {Gorti}}]{Adams:2004}
{Adams}, F.~C., {Hollenbach}, D., {Laughlin}, G., \& {Gorti}, U. 2004, \apj, 611, 360, \dodoi{10.1086/421989}

\bibitem[{{Alexander} {et~al.}(2014){Alexander}, {Pascucci}, {Andrews}, {Armitage}, \& {Cieza}}]{Alexander:2014}
{Alexander}, R., {Pascucci}, I., {Andrews}, S., {Armitage}, P., \& {Cieza}, L. 2014, in Protostars and Planets VI, ed. H.~{Beuther}, R.~S. {Klessen}, C.~P. {Dullemond}, \& T.~{Henning}, 475, \dodoi{10.2458/azu_uapress_9780816531240-ch021}

\bibitem[{{Anderson} {et~al.}(2013){Anderson}, {Adams}, \& {Calvet}}]{Anderson:2013}
{Anderson}, K.~R., {Adams}, F.~C., \& {Calvet}, N. 2013, \apj, 774, 9, \dodoi{10.1088/0004-637X/774/1/9}

\bibitem[{{Andrews}(2020)}]{Andrews:2020}
{Andrews}, S.~M. 2020, \araa, 58, 483, \dodoi{10.1146/annurev-astro-031220-010302}

\bibitem[{{Andrews} {et~al.}(2018){Andrews}, {Terrell}, {Tripathi}, {Ansdell}, {Williams}, \& {Wilner}}]{Andrews:2018}
{Andrews}, S.~M., {Terrell}, M., {Tripathi}, A., {et~al.} 2018, \apj, 865, 157, \dodoi{10.3847/1538-4357/aadd9f}

\bibitem[{{Ansdell} {et~al.}(2016){Ansdell}, {Williams}, {van der Marel}, {Carpenter}, {Guidi}, {Hogerheijde}, {Mathews}, {Manara}, {Miotello}, {Natta}, {Oliveira}, {Tazzari}, {Testi}, {van Dishoeck}, \& {van Terwisga}}]{Ansdell:2016}
{Ansdell}, M., {Williams}, J.~P., {van der Marel}, N., {et~al.} 2016, \apj, 828, 46, \dodoi{10.3847/0004-637X/828/1/46}

\bibitem[{{Ansdell} {et~al.}(2018){Ansdell}, {Williams}, {Trapman}, {van Terwisga}, {Facchini}, {Manara}, {van der Marel}, {Miotello}, {Tazzari}, {Hogerheijde}, {Guidi}, {Testi}, \& {van Dishoeck}}]{Ansdell:2018}
{Ansdell}, M., {Williams}, J.~P., {Trapman}, L., {et~al.} 2018, \apj, 859, 21, \dodoi{10.3847/1538-4357/aab890}

\bibitem[{{Bayo} {et~al.}(2012){Bayo}, {Barrado}, {Hu{\'e}lamo}, {Morales-Calder{\'o}n}, {Melo}, {Stauffer}, \& {Stelzer}}]{Bayo:2012}
{Bayo}, A., {Barrado}, D., {Hu{\'e}lamo}, N., {et~al.} 2012, \aap, 547, A80, \dodoi{10.1051/0004-6361/201219374}

\bibitem[{{Calvet} \& {Gullbring}(1998)}]{Calvet:1998}
{Calvet}, N., \& {Gullbring}, E. 1998, \apj, 509, 802, \dodoi{10.1086/306527}

\bibitem[{{Calvet} {et~al.}(2004){Calvet}, {Muzerolle}, {Brice{\~n}o}, {Hern{\'a}ndez}, {Hartmann}, {Saucedo}, \& {Gordon}}]{Calvet:2004}
{Calvet}, N., {Muzerolle}, J., {Brice{\~n}o}, C., {et~al.} 2004, \aj, 128, 1294, \dodoi{10.1086/422733}

\bibitem[{{Carpenter} {et~al.}(2006){Carpenter}, {Mamajek}, {Hillenbrand}, \& {Meyer}}]{Carpenter:2006}
{Carpenter}, J.~M., {Mamajek}, E.~E., {Hillenbrand}, L.~A., \& {Meyer}, M.~R. 2006, \apjl, 651, L49, \dodoi{10.1086/509121}

\bibitem[{{Clarke} {et~al.}(2001){Clarke}, {Gendrin}, \& {Sotomayor}}]{Clarke:2001}
{Clarke}, C.~J., {Gendrin}, A., \& {Sotomayor}, M. 2001, \mnras, 328, 485, \dodoi{10.1046/j.1365-8711.2001.04891.x}

\bibitem[{{de Gregorio-Monsalvo} {et~al.}(2013){de Gregorio-Monsalvo}, {M{\'e}nard}, {Dent}, {Pinte}, {L{\'o}pez}, {Klaassen}, {Hales}, {Cort{\'e}s}, {Rawlings}, {Tachihara}, {Testi}, {Takahashi}, {Chapillon}, {Mathews}, {Juhasz}, {Akiyama}, {Higuchi}, {Saito}, {Nyman}, {Phillips}, {Rod{\'o}n}, {Corder}, \& {Van Kempen}}]{DeGregorioMonsalvo:2013}
{de Gregorio-Monsalvo}, I., {M{\'e}nard}, F., {Dent}, W., {et~al.} 2013, \aap, 557, A133, \dodoi{10.1051/0004-6361/201321603}

\bibitem[{{Eisner} {et~al.}(2018){Eisner}, {Arce}, {Ballering}, {Bally}, {Andrews}, {Boyden}, {Di Francesco}, {Fang}, {Johnstone}, {Kim}, {Mann}, {Matthews}, {Pascucci}, {Ricci}, {Sheehan}, \& {Williams}}]{Eisner:2018}
{Eisner}, J.~A., {Arce}, H.~G., {Ballering}, N.~P., {et~al.} 2018, \apj, 860, 77, \dodoi{10.3847/1538-4357/aac3e2}

\bibitem[{{Facchini} {et~al.}(2017){Facchini}, {Birnstiel}, {Bruderer}, \& {van Dishoeck}}]{Facchini:2017}
{Facchini}, S., {Birnstiel}, T., {Bruderer}, S., \& {van Dishoeck}, E.~F. 2017, \aap, 605, A16, \dodoi{10.1051/0004-6361/201630329}

\bibitem[{{Feiden}(2016)}]{Feiden:2016}
{Feiden}, G.~A. 2016, \aap, 593, A99, \dodoi{10.1051/0004-6361/201527613}

\bibitem[{{Flaccomio} {et~al.}(2003){Flaccomio}, {Damiani}, {Micela}, {Sciortino}, {Harnden}, {Murray}, \& {Wolk}}]{Flaccomio:2003}
{Flaccomio}, E., {Damiani}, F., {Micela}, G., {et~al.} 2003, \apj, 582, 398, \dodoi{10.1086/344536}

\bibitem[{{Gorti} \& {Hollenbach}(2009)}]{Gorti:2009}
{Gorti}, U., \& {Hollenbach}, D. 2009, \apj, 690, 1539, \dodoi{10.1088/0004-637X/690/2/1539}

\bibitem[{{Gorti} {et~al.}(2015){Gorti}, {Hollenbach}, \& {Dullemond}}]{Gorti:2015}
{Gorti}, U., {Hollenbach}, D., \& {Dullemond}, C.~P. 2015, \apj, 804, 29, \dodoi{10.1088/0004-637X/804/1/29}

\bibitem[{{Grant} {et~al.}(2018){Grant}, {Espaillat}, {Megeath}, {Calvet}, {Fischer}, {Miller}, {Kim}, {Stutz}, {Ribas}, \& {Robinson}}]{Grant:2018}
{Grant}, S.~L., {Espaillat}, C.~C., {Megeath}, S.~T., {et~al.} 2018, \apj, 863, 13, \dodoi{10.3847/1538-4357/aacda7}

\bibitem[{{Haisch} {et~al.}(2001){Haisch}, {Lada}, \& {Lada}}]{Haisch:2001}
{Haisch}, Karl~E., J., {Lada}, E.~A., \& {Lada}, C.~J. 2001, \apjl, 553, L153, \dodoi{10.1086/320685}

\bibitem[{{Haworth} \& {Clarke}(2019)}]{Haworth:2019}
{Haworth}, T.~J., \& {Clarke}, C.~J. 2019, \mnras, 485, 3895, \dodoi{10.1093/mnras/stz706}

\bibitem[{{Hayashi}(1981)}]{Hayashi:1981}
{Hayashi}, C. 1981, Progress of Theoretical Physics Supplement, 70, 35, \dodoi{10.1143/PTPS.70.35}

\bibitem[{{Hendler} {et~al.}(2020){Hendler}, {Pascucci}, {Pinilla}, {Tazzari}, {Carpenter}, {Malhotra}, \& {Testi}}]{Hendler:2020}
{Hendler}, N., {Pascucci}, I., {Pinilla}, P., {et~al.} 2020, \apj, 895, 126, \dodoi{10.3847/1538-4357/ab70ba}

\bibitem[{{Hern{\'a}ndez} {et~al.}(2007){Hern{\'a}ndez}, {Hartmann}, {Megeath}, {Gutermuth}, {Muzerolle}, {Calvet}, {Vivas}, {Brice{\~n}o}, {Allen}, {Stauffer}, {Young}, \& {Fazio}}]{Hernandez:2007}
{Hern{\'a}ndez}, J., {Hartmann}, L., {Megeath}, T., {et~al.} 2007, \apj, 662, 1067, \dodoi{10.1086/513735}

\bibitem[{{Hollenbach} {et~al.}(1994){Hollenbach}, {Johnstone}, {Lizano}, \& {Shu}}]{Hollenbach:1994}
{Hollenbach}, D., {Johnstone}, D., {Lizano}, S., \& {Shu}, F. 1994, \apj, 428, 654, \dodoi{10.1086/174276}

\bibitem[{{Husser} {et~al.}(2013){Husser}, {Wende-von Berg}, {Dreizler}, {Homeier}, {Reiners}, {Barman}, \& {Hauschildt}}]{Husser:2013}
{Husser}, T.~O., {Wende-von Berg}, S., {Dreizler}, S., {et~al.} 2013, \aap, 553, A6, \dodoi{10.1051/0004-6361/201219058}

\bibitem[{{Hutchison} {et~al.}(2016){Hutchison}, {Laibe}, \& {Maddison}}]{Hutchison:2016}
{Hutchison}, M.~A., {Laibe}, G., \& {Maddison}, S.~T. 2016, \mnras, 463, 2725, \dodoi{10.1093/mnras/stw2191}

\bibitem[{{Kimura} {et~al.}(2016){Kimura}, {Kunitomo}, \& {Takahashi}}]{Kimura:2016}
{Kimura}, S.~S., {Kunitomo}, M., \& {Takahashi}, S.~Z. 2016, \mnras, 461, 2257, \dodoi{10.1093/mnras/stw1531}

\bibitem[{{Komaki} {et~al.}(2023){Komaki}, {Fukuhara}, {Suzuki}, \& {Yoshida}}]{Komaki:2023}
{Komaki}, A., {Fukuhara}, S., {Suzuki}, T.~K., \& {Yoshida}, N. 2023, arXiv e-prints, arXiv:2304.13316, \dodoi{10.48550/arXiv.2304.13316}

\bibitem[{{Komaki} {et~al.}(2024){Komaki}, {Kuiper}, \& {Yoshida}}]{Komaki:2024}
{Komaki}, A., {Kuiper}, R., \& {Yoshida}, N. 2024, \apj, 963, 81, \dodoi{10.3847/1538-4357/ad21f1}

\bibitem[{{Komaki} {et~al.}(2021){Komaki}, {Nakatani}, \& {Yoshida}}]{Komaki:2021}
{Komaki}, A., {Nakatani}, R., \& {Yoshida}, N. 2021, \apj, 910, 51, \dodoi{10.3847/1538-4357/abe2af}

\bibitem[{{Kroupa}(2001)}]{Kroupa:2001}
{Kroupa}, P. 2001, \mnras, 322, 231, \dodoi{10.1046/j.1365-8711.2001.04022.x}

\bibitem[{{Kunitomo} {et~al.}(2021){Kunitomo}, {Ida}, {Takeuchi}, {Pani{\'c}}, {Miley}, \& {Suzuki}}]{Kunitomo:2021}
{Kunitomo}, M., {Ida}, S., {Takeuchi}, T., {et~al.} 2021, \apj, 909, 109, \dodoi{10.3847/1538-4357/abdb2a}

\bibitem[{{Kunitomo} {et~al.}(2020){Kunitomo}, {Suzuki}, \& {Inutsuka}}]{Kunitomo:2020}
{Kunitomo}, M., {Suzuki}, T.~K., \& {Inutsuka}, S.-i. 2020, \mnras, 492, 3849, \dodoi{10.1093/mnras/staa087}

\bibitem[{{Mamajek}(2009)}]{Mamajek:2009}
{Mamajek}, E.~E. 2009, in American Institute of Physics Conference Series, Vol. 1158, American Institute of Physics Conference Series, ed. T.~{Usuda}, M.~{Tamura}, \& M.~{Ishii}, 3--10, \dodoi{10.1063/1.3215910}

\bibitem[{{Manara} {et~al.}(2023){Manara}, {Ansdell}, {Rosotti}, {Hughes}, {Armitage}, {Lodato}, \& {Williams}}]{Manara:2023}
{Manara}, C.~F., {Ansdell}, M., {Rosotti}, G.~P., {et~al.} 2023, in Astronomical Society of the Pacific Conference Series, Vol. 534, Protostars and Planets VII, ed. S.~{Inutsuka}, Y.~{Aikawa}, T.~{Muto}, K.~{Tomida}, \& M.~{Tamura}, 539, \dodoi{10.48550/arXiv.2203.09930}

\bibitem[{{Meyer} {et~al.}(2007){Meyer}, {Backman}, {Weinberger}, \& {Wyatt}}]{Meyer:2007}
{Meyer}, M.~R., {Backman}, D.~E., {Weinberger}, A.~J., \& {Wyatt}, M.~C. 2007, in Protostars and Planets V, ed. B.~{Reipurth}, D.~{Jewitt}, \& K.~{Keil}, 573.
\newblock \doarXiv{astro-ph/0606399}

\bibitem[{{Michel} {et~al.}(2021){Michel}, {van der Marel}, \& {Matthews}}]{Michel:2021}
{Michel}, A., {van der Marel}, N., \& {Matthews}, B.~C. 2021, \apj, 921, 72, \dodoi{10.3847/1538-4357/ac1bbb}

\bibitem[{{Muzerolle} {et~al.}(2003){Muzerolle}, {Hillenbrand}, {Calvet}, {Brice{\~n}o}, \& {Hartmann}}]{Muzerolle:2003}
{Muzerolle}, J., {Hillenbrand}, L., {Calvet}, N., {Brice{\~n}o}, C., \& {Hartmann}, L. 2003, \apj, 592, 266, \dodoi{10.1086/375704}

\bibitem[{{Nakatani} {et~al.}(2018){Nakatani}, {Hosokawa}, {Yoshida}, {Nomura}, \& {Kuiper}}]{Nakatani:2018}
{Nakatani}, R., {Hosokawa}, T., {Yoshida}, N., {Nomura}, H., \& {Kuiper}, R. 2018, \apj, 857, 57, \dodoi{10.3847/1538-4357/aab70b}

\bibitem[{{Owen} {et~al.}(2010){Owen}, {Ercolano}, {Clarke}, \& {Alexand er}}]{Owen:2010}
{Owen}, J.~E., {Ercolano}, B., {Clarke}, C.~J., \& {Alexand er}, R.~D. 2010, \mnras, 401, 1415, \dodoi{10.1111/j.1365-2966.2009.15771.x}

\bibitem[{{Pascucci} {et~al.}(2023){Pascucci}, {Cabrit}, {Edwards}, {Gorti}, {Gressel}, \& {Suzuki}}]{Pascucci:2023}
{Pascucci}, I., {Cabrit}, S., {Edwards}, S., {et~al.} 2023, in Astronomical Society of the Pacific Conference Series, Vol. 534, Protostars and Planets VII, ed. S.~{Inutsuka}, Y.~{Aikawa}, T.~{Muto}, K.~{Tomida}, \& M.~{Tamura}, 567, \dodoi{10.48550/arXiv.2203.10068}

\bibitem[{{Pascucci} {et~al.}(2016){Pascucci}, {Testi}, {Herczeg}, {Long}, {Manara}, {Hendler}, {Mulders}, {Krijt}, {Ciesla}, {Henning}, {Mohanty}, {Drabek-Maunder}, {Apai}, {Sz{\H{u}}cs}, {Sacco}, \& {Olofsson}}]{Pascucci:2016}
{Pascucci}, I., {Testi}, L., {Herczeg}, G.~J., {et~al.} 2016, \apj, 831, 125, \dodoi{10.3847/0004-637X/831/2/125}

\bibitem[{{Pascucci} {et~al.}(2020){Pascucci}, {Banzatti}, {Gorti}, {Fang}, {Pontoppidan}, {Alexander}, {Ballabio}, {Edwards}, {Salyk}, {Sacco}, {Flaccomio}, {Blake}, {Carmona}, {Hall}, {Kamp}, {K{\"a}ufl}, {Meeus}, {Meyer}, {Pauly}, {Steendam}, \& {Sterzik}}]{Pascucci:2020}
{Pascucci}, I., {Banzatti}, A., {Gorti}, U., {et~al.} 2020, \apj, 903, 78, \dodoi{10.3847/1538-4357/abba3c}

\bibitem[{{Paxton} {et~al.}(2011){Paxton}, {Bildsten}, {Dotter}, {Herwig}, {Lesaffre}, \& {Timmes}}]{Paxton:2011}
{Paxton}, B., {Bildsten}, L., {Dotter}, A., {et~al.} 2011, \apjs, 192, 3, \dodoi{10.1088/0067-0049/192/1/3}

\bibitem[{{Pfalzner} {et~al.}(2022){Pfalzner}, {Dehghani}, \& {Michel}}]{Pfalzner:2022}
{Pfalzner}, S., {Dehghani}, S., \& {Michel}, A. 2022, \apjl, 939, L10, \dodoi{10.3847/2041-8213/ac9839}

\bibitem[{{Picogna} {et~al.}(2019){Picogna}, {Ercolano}, {Owen}, \& {Weber}}]{Picogna:2019}
{Picogna}, G., {Ercolano}, B., {Owen}, J.~E., \& {Weber}, M.~L. 2019, \mnras, 487, 691, \dodoi{10.1093/mnras/stz1166}

\bibitem[{{Ribas} {et~al.}(2015){Ribas}, {Bouy}, \& {Mer{\'\i}n}}]{Ribas:2015}
{Ribas}, {\'A}., {Bouy}, H., \& {Mer{\'\i}n}, B. 2015, \aap, 576, A52, \dodoi{10.1051/0004-6361/201424846}

\bibitem[{{Ribas} {et~al.}(2014){Ribas}, {Mer{\'\i}n}, {Bouy}, \& {Maud}}]{Ribas:2014}
{Ribas}, {\'A}., {Mer{\'\i}n}, B., {Bouy}, H., \& {Maud}, L.~T. 2014, \aap, 561, A54, \dodoi{10.1051/0004-6361/201322597}

\bibitem[{{Richert} {et~al.}(2018){Richert}, {Getman}, {Feigelson}, {Kuhn}, {Broos}, {Povich}, {Bate}, \& {Garmire}}]{Richert:2018}
{Richert}, A.~J.~W., {Getman}, K.~V., {Feigelson}, E.~D., {et~al.} 2018, \mnras, 477, 5191, \dodoi{10.1093/mnras/sty949}

\bibitem[{{Ronco} {et~al.}(2024){Ronco}, {Schreiber}, {Villaver}, {Guilera}, \& {Miller Bertolami}}]{Ronco:2023}
{Ronco}, M.~P., {Schreiber}, M.~R., {Villaver}, E., {Guilera}, O.~M., \& {Miller Bertolami}, M.~M. 2024, \aap, 682, A155, \dodoi{10.1051/0004-6361/202347762}

\bibitem[{{Ru{\'\i}z-Rodr{\'\i}guez} {et~al.}(2018){Ru{\'\i}z-Rodr{\'\i}guez}, {Cieza}, {Williams}, {Andrews}, {Principe}, {Caceres}, {Canovas}, {Casassus}, {Schreiber}, \& {Kastner}}]{RuizRodriguez:2018}
{Ru{\'\i}z-Rodr{\'\i}guez}, D., {Cieza}, L.~A., {Williams}, J.~P., {et~al.} 2018, \mnras, 478, 3674, \dodoi{10.1093/mnras/sty1351}

\bibitem[{{Shu} {et~al.}(1994){Shu}, {Najita}, {Ostriker}, {Wilkin}, {Ruden}, \& {Lizano}}]{Shu:1994}
{Shu}, F., {Najita}, J., {Ostriker}, E., {et~al.} 1994, \apj, 429, 781, \dodoi{10.1086/174363}

\bibitem[{{Suzuki} {et~al.}(2016){Suzuki}, {Ogihara}, {Morbidelli}, {Crida}, \& {Guillot}}]{Suzuki:2016}
{Suzuki}, T.~K., {Ogihara}, M., {Morbidelli}, A., {Crida}, A., \& {Guillot}, T. 2016, \aap, 596, A74, \dodoi{10.1051/0004-6361/201628955}

\bibitem[{{Tanaka} {et~al.}(2013){Tanaka}, {Nakamoto}, \& {Omukai}}]{Tanaka:2013}
{Tanaka}, K. E.~I., {Nakamoto}, T., \& {Omukai}, K. 2013, \apj, 773, 155, \dodoi{10.1088/0004-637X/773/2/155}

\bibitem[{{Tobin} {et~al.}(2020){Tobin}, {Sheehan}, {Megeath}, {D{\'\i}az-Rodr{\'\i}guez}, {Offner}, {Murillo}, {van 't Hoff}, {van Dishoeck}, {Osorio}, {Anglada}, {Furlan}, {Stutz}, {Reynolds}, {Karnath}, {Fischer}, {Persson}, {Looney}, {Li}, {Stephens}, {Chandler}, {Cox}, {Dunham}, {Tychoniec}, {Kama}, {Kratter}, {Kounkel}, {Mazur}, {Maud}, {Patel}, {Perez}, {Sadavoy}, {Segura-Cox}, {Sharma}, {Stephenson}, {Watson}, \& {Wyrowski}}]{Tobin:2020}
{Tobin}, J.~J., {Sheehan}, P.~D., {Megeath}, S.~T., {et~al.} 2020, \apj, 890, 130, \dodoi{10.3847/1538-4357/ab6f64}

\bibitem[{{Tychoniec} {et~al.}(2018){Tychoniec}, {Tobin}, {Karska}, {Chandler}, {Dunham}, {Harris}, {Kratter}, {Li}, {Looney}, {Melis}, {P{\'e}rez}, {Sadavoy}, {Segura-Cox}, \& {van Dishoeck}}]{Tychoniec:2018}
{Tychoniec}, {\L}., {Tobin}, J.~J., {Karska}, A., {et~al.} 2018, \apjs, 238, 19, \dodoi{10.3847/1538-4365/aaceae}

\bibitem[{{Valenti} {et~al.}(2003){Valenti}, {Fallon}, \& {Johns-Krull}}]{Valenti:2003}
{Valenti}, J.~A., {Fallon}, A.~A., \& {Johns-Krull}, C.~M. 2003, \apjs, 147, 305, \dodoi{10.1086/375445}

\bibitem[{{Wang} \& {Goodman}(2017)}]{Wang:2017}
{Wang}, L., \& {Goodman}, J. 2017, \apj, 847, 11, \dodoi{10.3847/1538-4357/aa8726}

\bibitem[{{Weder} {et~al.}(2023){Weder}, {Mordasini}, \& {Emsenhuber}}]{Weder:2023}
{Weder}, J., {Mordasini}, C., \& {Emsenhuber}, A. 2023, \aap, 674, A165, \dodoi{10.1051/0004-6361/202243453}

\bibitem[{{Yasui} {et~al.}(2014){Yasui}, {Kobayashi}, {Tokunaga}, \& {Saito}}]{Yasui:2014}
{Yasui}, C., {Kobayashi}, N., {Tokunaga}, A.~T., \& {Saito}, M. 2014, \mnras, 442, 2543, \dodoi{10.1093/mnras/stu1013}

\end{thebibliography}
\bibliographystyle{aasjournal}



\end{document}